\documentclass[
reprint,
amsmath,amssymb,
aps,
prb,
twocolumn,
superscriptaddress,
nolongbibliography
]{revtex4-2}

\usepackage{graphicx}% Include figure files
\usepackage{dcolumn}% Align table columns on a decimal point
\usepackage{bm}% bold math
\usepackage{hyperref}% add hypertext capabilities
\usepackage{physics}
\usepackage{mhchem}

\begin{document}

\preprint{APS/123-QED}

\title{Four-Spin Interactions as a Route to Multiple-$Q$ Topological Magnetic Order}

\author{Kazuki Okigami}
\affiliation{Department of Applied Physics, The University of Tokyo, Bunkyo, Tokyo 113-8656, Japan}
\email{okigami@g.ecc.u-tokyo.ac.jp}
\author{Satoru Hayami}
\affiliation{Graduate school of Science, Hokkaido University, Sapporo 060-0810, Japan}
\email{hayami@phys.sci.hookudai.ac.jp}

\date{\today}

\begin{abstract}
We investigate the role of four-spin interactions in stabilizing exotic multiple-$Q$ topological spin textures and demonstrate their ability to realize a skyrmion crystal.
While such higher-order interactions are known to be important, their intricate nature makes systematic model construction significantly challenging.
To address this issue, we develop a theoretical framework that connects microscopic real-space four-spin couplings to their effective interactions in momentum space, providing a clear route to engineer target magnetic phases. 
Applying this framework to a frustrated Heisenberg model with designed four-spin interactions, we identify the stabilization of the zero-field skyrmion crystal with a topological number of two via simulated annealing. 
Furthermore, our momentum-space analysis reveals the intrinsic mechanism by which the well-known ring-exchange interaction also favors the skyrmion crystal. 
Our findings not only present a concrete model for a higher-order skyrmion crystal but also offer a general methodology for understanding and designing a wide range of complex multiple-$Q$ magnetic orders driven by multi-spin interactions.
\end{abstract}

\maketitle

\section{Introduction}
Spin textures in magnetic materials have garnered significant interest in condensed matter physics as they exhibit rich physical phenomena, including emergent electromagnetic fields, magnetoelectric effects, and anomalous transport properties~\cite{nagaosa2013topological,Tokura_Kanazawa_ChemRev.121.2857}. 
Among various spin textures, topologically non-trivial spin textures are in the center of focus since their topological nature makes them robust information carriers for device applications, such as racetrack memory and logic devices~\cite{fert2013skyrmions,zhang2020skyrmion,Psaroudaki2021SkyrmionQubit}.
Following the first experimental observation of periodic arrangements of skyrmions, i.e., a skyrmion crystal (SkX) in \ce{MnSi}~\cite{Muhlbauer_2009skyrmion}, the search for novel functionalities has driven the discovery of other tribes of two-dimensional topological spin textures~\cite{gobel2021beyond, Zhou2025-topological}. 
These include experimentally observed skyrmioniums~\cite{Zhang2018Skyrmionium}, and antiferromagnetic SkX~\cite{Gao2016Spiral, gao2020fractional}. 
Recently, three-dimensional topological spin textures, such as hedgehogs~\cite{fujishiro2019topological,Kanazawa_PhysRevLett.125.137202,Ishiwata_PhysRevB.101.134406} and hopfions~\cite{Kent2021-hopifon, Yu_AdvMater.202210646_Hopfion,zheng2023hopfion}, have also been discovered, which has further expanded the research field of topological spin textures.
Among them, skyrmions have been extensively studied due to their stability against perturbations and intriguing transport phenomena that could be applied in devices, such as the topological Hall effect~\cite{Ohgushi_PhysRevB.62.R6065, Neubauer_PhysRevLett.102.186602, Hamamoto_PhysRevB.92.115417, kurumaji2019skyrmion} and the topological Nernst effect~\cite{Shiomi_PhysRevB.88.064409, Hirschberger_Nernst_prl2020}.
To realize the development of %skyrmion-based 
devices based on topological spin textures, it is crucial to establish stabilization mechanisms for various types of topological spin textures.

Theoretical studies have proposed various mechanisms for stabilizing topological magnetic orders including SkXs~\cite{HayamiYambe2024_stabilization}, such as the Dzyaloshinskii-Moriya (DM) interaction~\cite{dzyaloshinsky1958thermodynamic,moriya1960anisotropic} in non-centrosymmetric magnets~\cite{rossler2006spontaneous, Yi_PhysRevB.80.054416}, magnetic frustration in centrosymmetric magnets~\cite{Okubo_PhysRevLett.108.017206,leonov2015multiply, Lin_PhysRevB.93.064430}, dipole-dipole interactions~\cite{Ezawa_PhysRevLett.105.197202,Utesov_PhysRevB.103.064414, Utesov_PhysRevB.105.054435}, magnetic anisotropic interactions~\cite{amoroso2020spontaneous, Hayami_PhysRevB.103.054422, yambe2021skyrmion, amoroso2021tuning}, and long-range interactions mediated by itinerant electrons~\cite{Martin_PhysRevLett.101.156402,heinze2011spontaneous,takagi2018multiple, Ozawa_PhysRevLett.118.147205,Hayami_PhysRevB.95.224424, Nikolic_PhysRevB.103.155151}. 
Beyond these bilinear models, recent studies have revealed that higher-order spin interactions, such as three-spin~\cite{Mankovsky_PhysRevB.101.174401}, four-spin% interactions
~\cite{MacDonald_PhysRevB.37.9753_t/U, ueland2012controllable, Hoffmann_PhysRevB.101.024418}, and six-spin interactions~\cite{grytsiuk2020topological, Bomerich_PhysRevB.102.100408, hayami2021phase} play a significant role in stabilizing SkXs~\cite{Hayami_PhysRevB.95.224424, Paul2020role}.

Such multi-spin interactions are not merely phenomenological but arise from microscopic electronic mechanisms. 
They can emerge from strong electron correlation effects, as demonstrated through perturbative expansions of fundamental models, such as the Kondo lattice model~\cite{Akagi_PhysRevLett.108.096401, Hayami_PhysRevB.90.060402, Hayami_PhysRevB.95.224424} and the Hubbard model~\cite{takahashi1977half, yoshimori1978fourth, MacDonald_PhysRevB.37.9753_t/U, Hoffmann_PhysRevB.101.024418}. 
Reflecting their physical importance, the contributions of such four-spin interactions have also been quantitatively evaluated in various materials~\cite{Kartsev2020-biquadratic,Ni2021-biquadratic,Kirstein2025-fourspin}. 
In parallel, advanced first principles calculations have been developed to quantify these higher-order interactions by parameterizing effective spin Hamiltonians, including four-spin terms~\cite{Simon2020-biquadratic, Jacobsson2022_SciRep,Hatanaka2024-biquadratic}. 

Furthermore, machine learning techniques have been employed to construct effective spin models by learning from the energy landscapes and spin configurations of fundamental electronic models~\cite{fujita2018constructing,li2020constructing,sharma2023klm}.
In a different direction, recent efforts have focused on the inverse problem of constructing models that stabilize a target spin texture. 
Our recent work, for instance, demonstrated a successful inverse design of bilinear interaction models to stabilize specific SkXs by optimizing their momentum-space representation~\cite{Okigami2024-cp}.
This suggests that such a design principle could be a powerful tool, yet its extension to systematically construct models with higher-order interactions remains a significant challenge.

In this paper, directly extending our inverse design approach on bilinear interactions~\cite{Okigami2024-cp}, we explore the crucial role of four-spin interactions in stabilizing complex, topologically non-trivial spin textures.
Specifically, we demonstrate that the inclusion of these higher-order terms to the standard Heisenberg model is a powerful mechanism for realizing states such as the SkX structure with a skyrmion number of two (SkX2).
However, since the nature of these quartic couplings is far from trivial, judiciously parameterizing the relevant real-space couplings is challenging. 

We propose a methodology that leverages the momentum-space representation of the quartic interaction to systematically parametrize the necessary real-space four-spin couplings. 
Employing this technique, we first show, via simulated annealing (SA), that a carefully chosen set of simplified four-spin interactions successfully stabilizes the SkX2 phase. 
Furthermore, to connect our findings to established concepts, we investigate the influence of the well-known ring-exchange interaction~\cite{Thouless1965-ring, Momoi_PhysRevLett.79.2081}---a natural form of four-spin coupling---and explicitly demonstrate its ability to also induce and stabilize the SkX2 state. 

The remainder of this paper is organized as follows. In Sec.~II, we detail four-spin interactions and the momentum-space parametrization method. 
In Sec.~III, we present our numerical results of a model constructed using the inverse design approach. 
In Sec.~IV, we show the results for the ring-exchange model. 
Finally, in Sec.~V, we provide a discussion, followed by a concise summary in Sec.~VI.

\begin{figure*}[ht!]
  \begin{center}
  \includegraphics[width=0.999\hsize]{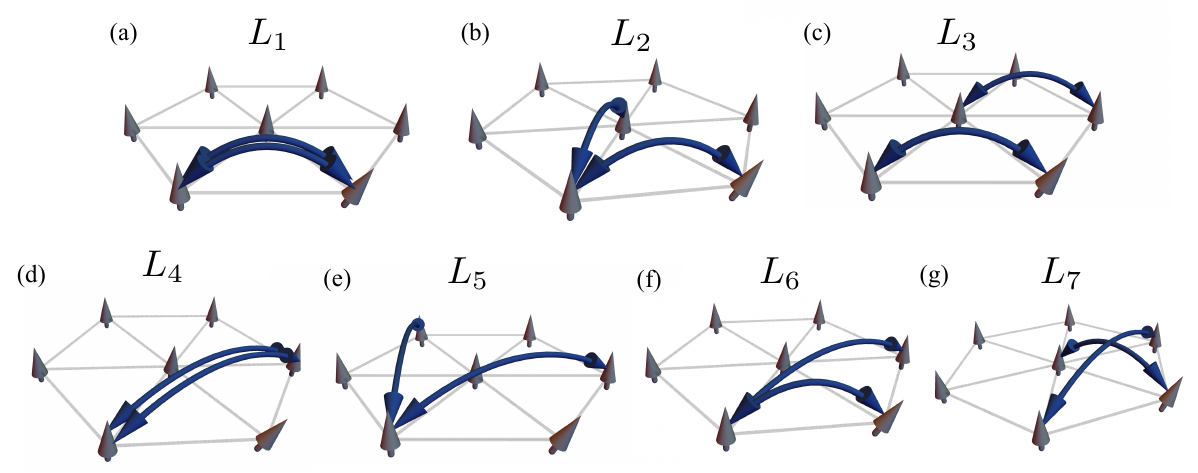}
  \caption{
%  (Color online)
  \label{fig: real-space four-body interactions}
  Schematic illustrations of the seven types of four-body interactions considered in this study, characterized by the coefficients $L_1$ to $L_7$ from (a) to (g).
  }
  \end{center}
\end{figure*}

\section{Four-body interactions}

\subsection{General formulation of four-body interactions}

To account for higher-order exchange effects, we extend the conventional bilinear Heisenberg Hamiltonian by including a quartic interaction term, $H_4$. 
In the following, we focus on the case of a two-dimensional triangular lattice.
This term involves the product of four spins and is generally defined as: 
\begin{align}
    H_4 = \sum_{i,j,k,l} L_{ijkl} (\bm{S}_i \cdot \bm{S}_j)(\bm{S}_k \cdot \bm{S}_l).
\end{align}
The parameters $L_{ijkl}$, which will be simplified to a few independent couplings ($L_1,L_2,\ldots$) based on the lattice symmetry and interaction range, are crucial for describing complex magnetic phenomena such as biquadratic interactions~\cite{Kartsev2020-biquadratic, Ni2021-biquadratic, Hatanaka2024-biquadratic} and ring exchange~\cite{Thouless1965-ring, Momoi_PhysRevLett.79.2081} that are not captured by the standard Heisenberg model within the bilinear exchange interaction.

To analyze the contribution of the quartic coupling $H_4$ within the context of stabilizing complex spin textures, we transform the real-space spins into momentum space using the Fourier transform $\bm{S}_{\bm{q}} = \frac{1}{\sqrt{N}} \sum_{i} e^{-i\bm{q}\cdot\bm{r}_i} \bm{S}_i$, where $\bm{r}_i$ is the position vector of site $i$ and $N$ is the total number of lattice sites. 
The general form of the quartic interaction term in momentum space is given by:
\begin{align}
  H_{4} = \frac{1}{N} \sum_{\bm{q},\bm{q}',\bm{q}''} K_{\bm{q},\bm{q}',\bm{q}''} (\bm{S}_{\bm{q}} \cdot \bm{S}_{\bm{q}'}) (\bm{S}_{\bm{q}''} \cdot \bm{S}_{-\bm{q}-\bm{q}'-\bm{q}''}),
\end{align}
where 
\begin{equation}
  \begin{aligned}
    K_{\bm{q},\bm{q}',\bm{q}''} = &\sum_{\{jkl\}_i} L_{ijkl} \\
    &\times e^{-i(\bm{q}'\cdot(\bm{r}_j - \bm{r}_i) + \bm{q}''\cdot(\bm{r}_k - \bm{r}_i) - (\bm{q}+\bm{q}'+\bm{q}'')\cdot(\bm{r}_l - \bm{r}_i))}
    \label{eq:K_q_q'_q''}
    .
  \end{aligned}
\end{equation}
Here, $\{jkl\}_i$ represents the set of sites $j,k,l$ related to site $i$.
The term $\bm{S}_{-\bm{q}-\bm{q}'-\bm{q}''}$ explicitly ensures the conservation of crystal momentum, where the sum of all four wave vectors must be zero up to reciprocal lattice vectors. 
In the subsequent analysis, the general function $K_{\bm{q}, \bm{q}', \bm{q}''}$ will be parametrized into a simplified set of five independent coupling functions, $K_1,\ldots,K_5$, by considering specific interactions relevant to our interests~\cite{Hayami_PhysRevB.95.224424}.

\subsection{Real-space four-body interactions}

To concretely investigate the contribution of specific four-spin interactions, we introduce seven distinct real-space interaction terms, characterized by the coefficients $L_1$ to $L_7$~\cite{sharma2023klm}. 
The geometric arrangements of the spins involved in each term are schematically illustrated in Fig.~\ref{fig: real-space four-body interactions}.

These interactions can be classified into two groups based on the range of the constituent bonds. 
The first group, governed by the coefficients $L_1$, $L_2$, and $L_3$, is constructed exclusively from bonds between nearest-neighbor (NN) sites. 
These terms represent fundamental and local coupling configurations of two bilinear products, $(\bm{S}_i \cdot \bm{S}_j)$ and $(\bm{S}_k \cdot \bm{S}_l)$, on the elementary plaquettes of the lattice. 

To explore the role of longer-range correlations, the second group of terms: $L_4$, $L_5$, $L_6$, and $L_7$, incorporates interactions extending up to next-nearest-neighbor (NNN) sites. 
The inclusion of these longer-range terms allows for a richer variety of spin correlations, which can be crucial for stabilizing more complex and larger-period magnetic structures that are not easily induced by the former interactions alone. 
Although we restricted our analysis to these seven types of interactions for simplicity, the framework can be readily extended to include other four-spin terms as needed to capture more intricate spin structures. 

The total four-spin Hamiltonian considered in this work is thus given by a linear combination of these seven terms:
\begin{equation}
H_4 = \sum_{\alpha=1}^{7} L_{\alpha} \sum_{\langle i,j,k,l \rangle \in \alpha} (\bm{S}_i \cdot \bm{S}_j)(\bm{S}_k \cdot \bm{S}_l),
\label{eq:H4}
\end{equation}
where the inner sum $\sum_{\langle i,j,k,l \rangle \in \alpha}$ runs over all quartets of sites $\{i,j,k,l\}$ that form the geometric configuration of type $\alpha$, as defined in Fig.~\ref{fig: real-space four-body interactions}.

\begin{figure}[bt!]
  \begin{center}
  \includegraphics[width=0.99\hsize]{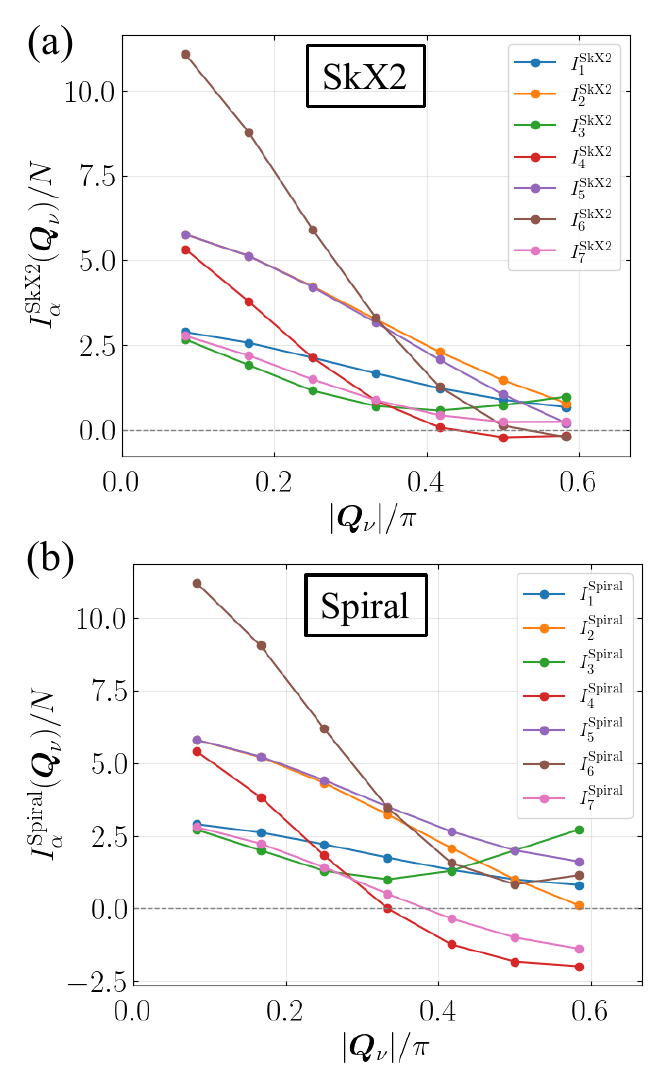}
  \caption{
%  (Color online)
  \label{fig: real-space energy}
  Energy contributions of each four-spin interaction term for (a) the SkX2 state and (b) the spiral state as a function of the ordering wave vector magnitude $|\bm{Q}_{\nu}|$. 
  }
  \end{center}
\end{figure}

\begin{figure}[bt!]
  \begin{center}
  \includegraphics[width=0.99\hsize]{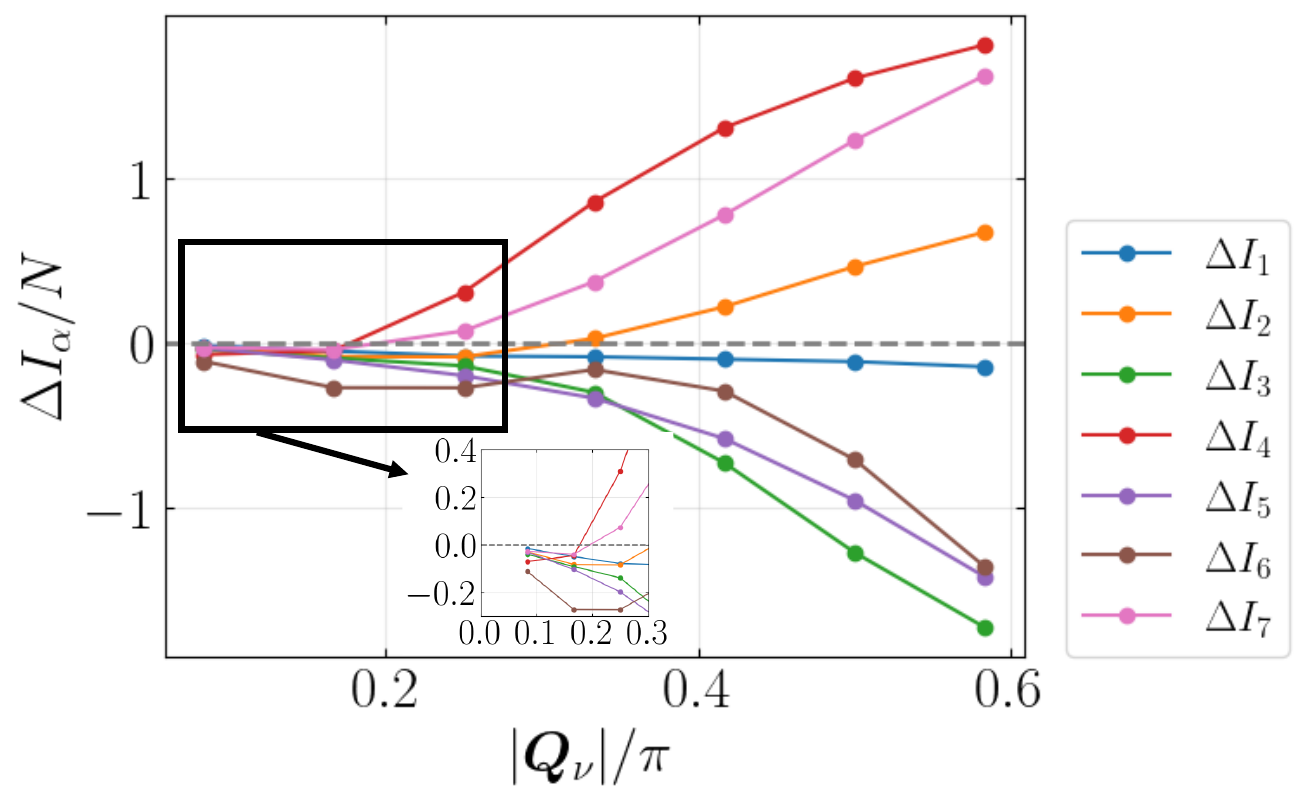}
  \caption{
%  (Color online)
  \label{fig: real-space energy difference}
  Energy contribution difference between the SkX2 and spiral states, $\Delta I_{\alpha} = I^{\text{SkX2}}_{\alpha} - I^{\text{Spiral}}_{\alpha}$, for each four-spin interaction term as a function of the ordering wave vector magnitude $|\bm{Q}_{\nu}|$.
  }
  \end{center}
\end{figure}

The objective of our study is to stabilize the SkX2 phase by introducing four-spin interactions from a model whose ground state is otherwise a simple spiral structure. 
The stabilization can be achieved if the four-spin interactions provide a substantial energy reduction for the SkX2 state relative to the spiral state. 
To investigate this, we evaluate the energy contribution of each four-spin term ($L_1$ to $L_7$) for these two competing magnetic structures. 

We employ the following ansatz for the spiral and SkX2 spin configurations on a triangular lattice~\cite{Ozawa_PhysRevLett.118.147205}:
\begin{align}
  \text{Spiral:}\quad \bm{S}_i &= (\cos(\bm{Q}_1 \cdot \bm{r}_i), \sin(\bm{Q}_1 \cdot \bm{r}_i), 0), \label{eq:ansatz_spiral} \\
  \text{SkX2:}\quad \bm{S}_i &= \mathcal{N} (\cos(\bm{Q}_1 \cdot \bm{r}_i), \cos(\bm{Q}_2 \cdot \bm{r}_i), \cos(\bm{Q}_3 \cdot \bm{r}_i)), \label{eq:ansatz_skx2}
\end{align}
where $\bm{Q}_\nu$ ($\nu=1,2,3$) are the symmetry-connected ordering wave vectors: $\bm{Q}_1 = |\bm{Q}_\nu|(1, 0), \bm{Q}_2 = |\bm{Q}_\nu|(-1/2, \sqrt{3}/2), \bm{Q}_3 = |\bm{Q}_\nu|(-1/2, -\sqrt{3}/2)$, and $\mathcal{N}$ is a normalization factor ensuring $|\bm{S}_i|=1$; we suppose that the ordering wave vectors lie on the $\langle 100 \rangle$ direction by considering appropriate bilinear exchange interactions. 
The SkX2 state is characterized by a superposition of three sinusoidal waves with equal amplitudes, which is in contrast to the conventional SkX with the topological number of one (SkX1) that are described by a superposition of three spiral waves.
The former SkX2 tends to be stabilized at zero field, while the latter SkX1 tends to be stabilized under an external magnetic field.

We then calculate the energetic contribution of each term $\alpha$ by defining the interaction energy per site for a given state $\Gamma \in \{\text{SkX2, Spiral}\}$ as
\begin{equation}
I^{\Gamma}_{\alpha} = \frac{1}{N} \sum_{\langle i,j,k,l \rangle \in \alpha} (\bm{S}_i \cdot \bm{S}_j)(\bm{S}_k \cdot \bm{S}_l).
\label{eq:I_beta_alpha}
\end{equation}
This quantity corresponds to the energy contribution of the $\alpha$-th term when its coupling constant is set to $L_\alpha=1$.

The calculated values of $I^{\text{SkX2}}_{\alpha}$ and $I^{\text{Spiral}}_{\alpha}$ are plotted as a function of the wave vector magnitude $|\bm{Q}_{\nu}|$ in Figs.~\ref{fig: real-space energy}(a) and (b), respectively.
For both states, we observe that most four-spin terms tend to be repulsive, which increases the total energy over a wide range of $|\bm{Q}_{\nu}|$. 
However, as $|\bm{Q}_{\nu}|$ increases, the magnitude of these contributions generally diminishes, and several terms become attractive, acting to lower the energy.

To directly compare the relative stability, we plot the energy difference, $\Delta I_{\alpha} = I^{\text{SkX2}}_{\alpha} - I^{\text{Spiral}}_{\alpha}$, in Fig.~\ref{fig: real-space energy difference}. 
The results reveal that some interactions always act to lower the energy of the SkX2 state relative to the spiral state (i.e., $\Delta I_{\alpha} < 0$). 
For other terms, the effect is sign-dependent on the value of $|\bm{Q}_{\nu}|$. 
This analysis confirms that an appropriate choice of a single four-spin interaction term and its magnitude can indeed render the SkX2 state energetically favorable; 
however, designing a model by combining multiple interactions to achieve a stable SkX2 phase remains a challenging task based on this analysis alone. 
A systematic approach is required to navigate the vast parameter space of the coupling constants $\{L_\alpha\}$.

\subsection{Momentum-space four-body interactions}

To systematically explore the parameter space of four-spin interactions and their influence on stabilizing complex spin textures, we change our perspective from real-space couplings $L_{\alpha}$ to momentum-space interactions $K_{\bm{q},\bm{q}',\bm{q}''}$. 
This transformation allows us to directly target the wave vectors associated with the desired magnetic structures, such as the SkX2, which is characterized by multiple ordering wave vectors. 
Therefore, in the general expression for $H_{4}$, we primarily consider contributions where the wave vectors $\{\bm{q}, \bm{q}', \bm{q}''\}$ are restricted to the set $\{\pm \bm{Q}_1, \pm \bm{Q}_2, \pm \bm{Q}_3\}$. 

This approach is motivated by the physical context of our model. 
We consider the four-spin Hamiltonian, $H_4$, as a perturbation to a parent Hamiltonian whose ground state is originally a spiral spin structure within the bilinear exchange interaction. 
Such spiral ground states under the bilinear exchange interaction are common in systems with the frustrated Heisenberg model and systems with competing ferromagnetic interactions and DM interactions. 
In these models on two-dimensional lattices, the ground state is typically degenerate among symmetry-equivalent ordering wave vectors, $\bm{Q}_\nu$, implying a possibility of their superpositions.
The primary role of the four-spin interaction is to induce the instability toward multiple-$Q$ states, which consists of a superposition of such symmetry-equivalent ordering wave vectors.
In this sense, our focus is naturally restricted to the interactions between the $\bm{S}_{\bm{q}}$ at these characteristic wave vectors, as they are the fundamental building blocks of the emergent complex magnetic order.

To simplify the situation, we omit the contribution from a finite uniform spin component, $\bm{S}_{\bm{0}}$, which arises from an external magnetic field or spontaneous magnetization, since our study specifically focuses on magnetic structures realized in the absence of an external magnetic field, where both spiral states and SkX2 exhibit $\bm{S}_{\bm{0}} = \bm{0}$.
In a general context, this uniform component, through the general four-spin interaction, can couple with the multiple-$Q$ modes, leading to complex terms such as $(\bm{S}_{\bm{0}} \cdot \bm{S}_{\bm{Q}_1}) (\bm{S}_{\bm{Q}_2} \cdot \bm{S}_{\bm{Q}_3})$ and their symmetry equivalents.
Thus, such contributions may play an important role in discussing magnetic instabilities under finite net magnetization; however, they are neglected in the present study.

When considering a multiple-$Q$ structure naturally stabilized on the underlying Heisenberg model, the only considerable Fourier components of the spin configuration are those at the characteristic wave vectors, $\{\bm{Q}_\nu\}$. 
This simplifies the general momentum-space Hamiltonian, as the sum of wave vectors $\bm{q}$, $\bm{q}'$, and $\bm{q}''$ in the generic interaction term $K_{\bm{q},\bm{q}',\bm{q}''} (\bm{S}_{\bm{q}} \cdot \bm{S}_{\bm{q}'}) (\bm{S}_{\bm{q}''} \cdot \bm{S}_{-\bm{q}-\bm{q}'-\bm{q}''})$ are restricted to this discrete set. 
Consequently, the vast number of possible interactions collapses into just five distinct functional forms, which we parameterize by the effective coupling constants $K_1$ through $K_5$. 
The allowed combinations of wave vectors for each of these five interaction types are summarized in Table~\ref{table: allowed set of wave vectors}.

\begin{table}[bt!]
  \caption{Allowed set of wave vectors for each $K$ term.}
  \label{table: allowed set of wave vectors}
  \centering
  \begin{tabular}{c|c||c|c|c|c|c|}
    \hline
      &  & $\bm{q}$ & $\bm{q}'$ & $\bm{q}''$  \\
    \hline \hline
    $K_1$  & $(\bm{S}_{\bm{Q}_{\nu}} \cdot \bm{S}_{-\bm{Q}_{\nu}})^2$ & $\bm{Q}_{\nu}$  & $-\bm{Q}_{\nu}$ & $\bm{Q}_{\nu}$ \\
      & & $\bm{Q}_{\nu}$  & $-\bm{Q}_{\nu}$ & $-\bm{Q}_{\nu}$ \\
    \hline
    $K_2$ & $(\bm{S}_{\bm{Q}_{\nu}} \cdot \bm{S}_{\bm{Q}_{\nu}}) (\bm{S}_{-\bm{Q}_{\nu}} \cdot \bm{S}_{-\bm{Q}_{\nu}})$ & $\bm{Q}_{\nu}$ & $\bm{Q}_{\nu}$ & $-\bm{Q}_{\nu}$ \\
    \hline
    $K_3$  & $(\bm{S}_{\bm{Q}_{\nu}} \cdot \bm{S}_{-\bm{Q}_{\nu}}) (\bm{S}_{\bm{Q}_{\nu'}} \cdot \bm{S}_{-\bm{Q}_{\nu'}})$ & $\bm{Q}_{\nu}$ & $-\bm{Q}_{\nu}$ & $\bm{Q}_{\nu'}$ \\
      & & $\bm{Q}_{\nu}$  & $-\bm{Q}_{\nu}$ & $-\bm{Q}_{\nu'}$ \\
    \hline
    $K_4$  & $(\bm{S}_{\bm{Q}_{\nu}} \cdot \bm{S}_{-\bm{Q}_{\nu'}}) (\bm{S}_{\bm{Q}_{\nu'}} \cdot \bm{S}_{-\bm{Q}_{\nu}})$ & $\bm{Q}_{\nu}$ & $-\bm{Q}_{\nu'}$ & $\bm{Q}_{\nu'}$ \\
      & & $\bm{Q}_{\nu}$  & $-\bm{Q}_{\nu'}$ & $-\bm{Q}_{\nu}$ \\
    \hline
    $K_5$  & $(\bm{S}_{\bm{Q}_{\nu}} \cdot \bm{S}_{\bm{Q}_{\nu'}}) (\bm{S}_{-\bm{Q}_{\nu}} \cdot \bm{S}_{-\bm{Q}_{\nu'}})$ & $\bm{Q}_{\nu}$ & $\bm{Q}_{\nu'}$ & $-\bm{Q}_{\nu}$ \\
      & & $\bm{Q}_{\nu}$  & $\bm{Q}_{\nu'}$ & $-\bm{Q}_{\nu'}$ \\
    \hline
  \end{tabular}
\end{table}

Based on these allowed interaction channels, the effective four-spin Hamiltonian for the multiple-$Q$ state, denoted as $H_4^K$, can be expressed as a sum of these five terms~\cite{Hayami_PhysRevB.95.224424}:
\begin{equation}
  \begin{aligned}
    H_4^K = \frac{1}{N} \sum_{\nu, \nu'} \bigg[ 
    & K_1 \sum_{\nu} \left( \bm{S}_{\bm{Q}_{\nu}} \cdot \bm{S}_{-\bm{Q}_{\nu}} \right)^2 \\
    &+ K_2 \sum_{\nu} \left( \bm{S}_{\bm{Q}_{\nu}} \cdot \bm{S}_{\bm{Q}_{\nu}} \right) \left( \bm{S}_{-\bm{Q}_{\nu}} \cdot \bm{S}_{-\bm{Q}_{\nu}} \right) \\
    &+ K_3 \sum_{\nu \neq \nu'} \left( \bm{S}_{\bm{Q}_{\nu}} \cdot \bm{S}_{-\bm{Q}_{\nu}} \right) \left( \bm{S}_{\bm{Q}_{\nu'}} \cdot \bm{S}_{-\bm{Q}_{\nu'}} \right) \\
    &+ K_4 \sum_{\nu \neq \nu'} \left( \bm{S}_{\bm{Q}_{\nu}} \cdot \bm{S}_{-\bm{Q}_{\nu'}} \right) \left( \bm{S}_{\bm{Q}_{\nu'}} \cdot \bm{S}_{-\bm{Q}_{\nu}} \right) \\
    &+ K_5 \sum_{\nu \neq \nu'} \left( \bm{S}_{\bm{Q}_{\nu}} \cdot \bm{S}_{\bm{Q}_{\nu'}} \right) \left( \bm{S}_{-\bm{Q}_{\nu}} \cdot \bm{S}_{-\bm{Q}_{\nu'}} \right)
    \bigg].
  \end{aligned}
  \label{eq:H4K}
\end{equation}

To understand the influence of each momentum-space interaction term ($K_1$ to $K_5$) on the stability of different magnetic states, we evaluate their energy contributions for both the single-$Q$ spiral and the SkX2 configurations using the ansatz defined in Eqs.~(\ref{eq:ansatz_spiral}) and (\ref{eq:ansatz_skx2}). 
Those ansatz are transformed into momentum space as follows:
\begin{align}
\text{Spiral:}\quad \bm{S}_{\bm{q}} &= \frac{\sqrt{N}}{2} (\delta_{\bm{q},\bm{Q}_1} + \delta_{\bm{q},-\bm{Q}_1}, -i(\delta_{\bm{q},\bm{Q}_1} - \delta_{\bm{q},-\bm{Q}_1}), 0), \\
\text{SkX2:}\quad \bm{S}_{\bm{q}} &= \sqrt{\frac{N}{6}} (\delta_{\bm{q},\bm{Q}_1} + \delta_{\bm{q},-\bm{Q}_1}, \nonumber \\
&\qquad\qquad \delta_{\bm{q},\bm{Q}_2} + \delta_{\bm{q},-\bm{Q}_2}, \nonumber \\
&\qquad\qquad \delta_{\bm{q},\bm{Q}_3} + \delta_{\bm{q},-\bm{Q}_3}).
\end{align}
Since SkX2 is a superposition of three sinusoidal waves, components of $\bm{S}_{\bm{Q}_{\nu}}$ for all $\nu$ are pure real numbers, while in the spiral state, $\bm{S}_{\bm{Q}_1}$ has both real and imaginary parts.

In the case of the spiral state, although $\bm{S}_{\bm{Q}_1} \cdot \bm{S}_{-\bm{Q}_1}$ is nonzero, $\bm{S}_{\bm{Q}_1} \cdot \bm{S}_{\bm{Q}_1} = 0$ holds, leading to the vanishing of the $K_2$ term.
On the other hand, for the SkX2 state, $\bm{S}_{\bm{Q}_{\nu}} \cdot \bm{S}_{\bm{Q}_{\nu}} \neq 0$ holds for all $\nu$, resulting in a nonzero contribution from the $K_2$ term.
Therefore, this $K_2$ term is expected to only contribute to the energy of the SkX2 state, potentially stabilizing it relative to the spiral state~\cite{hayami2024skyrmion}.
The $K_3$ term also solely contributes to the energy of the SkX2 state since it involves products of terms like $\bm{S}_{\bm{Q}_{\nu}} \cdot \bm{S}_{-\bm{Q}_{\nu}}$ for different $\nu$, which are nonzero in the SkX2 configuration but vanish in the single-$Q$ spiral state.
The other terms, $K_4$ and $K_5$, involve products of $\bm{S}_{\bm{q}}$ across different wave vectors, which vanish for both the spiral and SkX2 states.
The reason is that in the spiral state, only one wave vector $\bm{Q}_1$ is present, and in the SkX2 state, the orthogonality of the spin components at different $\bm{Q}_{\nu}$ ensures that these %cross 
terms do not contribute, which persists under the SO(3) rotational operation of the spin space.
The results of this evaluation are summarized in Table~\ref{table: effect of K}.

\begin{table}[tb!]
  \caption{Effect of momentum-space interactions $K$ on energies.}
  \label{table: effect of K}
  \centering
  \begin{tabular}{c||c|c|c|c|c|}
    \hline
    state  & $K_1$ & $K_2$ & $K_3$ & $K_4$ & $K_5$  \\
    \hline
    Spiral  & $+$ & 0  & 0 & 0 & 0 \\
    SkX2  & $+$  & $+$  & $+$ & 0 & 0 \\
    \hline
  \end{tabular}
\end{table}

Even though all nonzero contributions from the four-spin interactions are repulsive for positive $K_{\beta}$ for $\beta = 1,2,3$ (i.e., they increase the total energy), their action differs between the spiral and SkX2 states. 
We can analyze the energetic influence of each $K$ term on the spin structures as follows.

The biquadratic term, $K_1$, is repulsive for $K_1 > 0$. 
It energetically penalizes large single-mode amplitudes, which in turn promotes the distribution of spin components over multiple wave vectors and tends to stabilize multiple-$Q$ states~\cite{Hayami_PhysRevB.95.224424}. 
While this term raises the energy of both the SkX2 and spiral phases, the energy penalty is suppressed for the SkX2 state, where the spin components are already distributed among several $\bm{Q}_\nu$. 
Consequently, a positive $K_1$ effectively stabilizes the SkX2 state relative to the single-$Q$ spiral state.

The $K_2$ term provides a more selective stabilization mechanism. This interaction vanishes for a superposition of helical spirals but remains finite for a superposition of sinusoidal waves, which is the case for the SkX2 state. Therefore, a negative contribution ($K_2 < 0$) exclusively lowers the energy of the SkX2 state, providing a direct attractive stabilization.

Similarly, the $K_3$ term is naturally inactive for any single-$Q$ order such as a spiral, as it requires contributions from different wave vectors ($\nu \neq \nu'$). A negative $K_3$ thus favors the formation of a multiple-$Q$ state. However, unlike $K_2$, this term is also finite for other multiple-$Q$ structures, including those formed by superimposing helical spirals, as found in the SkX1. 
As a result, it does not selectively favor the sinusoidal SkX2 state over other possible multiple-$Q$ phases.

These analyses highlight the distinct roles that different four-spin interactions play in determining the relative stability of complex magnetic textures, providing insights into the design of effective spin models for realizing desired spin structures.

\subsection{Mapping from Real-Space to Momentum-Space Couplings}

To connect the four-spin interactions in real and momentum space microscopically, we express the effective momentum-space couplings $\{K_{\beta}\}$ in terms of each real-space coupling $L_\alpha$. 
This is achieved by performing the lattice summation in the general expression for the momentum-space interaction given in Eq.~(\ref{eq:K_q_q'_q''}). 
The contribution from a single real-space term $L_\alpha$ to this expression is denoted as $K_{\bm{q},\bm{q}',\bm{q}''}^{\alpha}$.
The explicit expressions are in the Appendix~\ref{appendix: derivation of K from L}.

\begin{figure*}[htb!]
  \begin{center}
    \includegraphics[width=0.99\hsize]{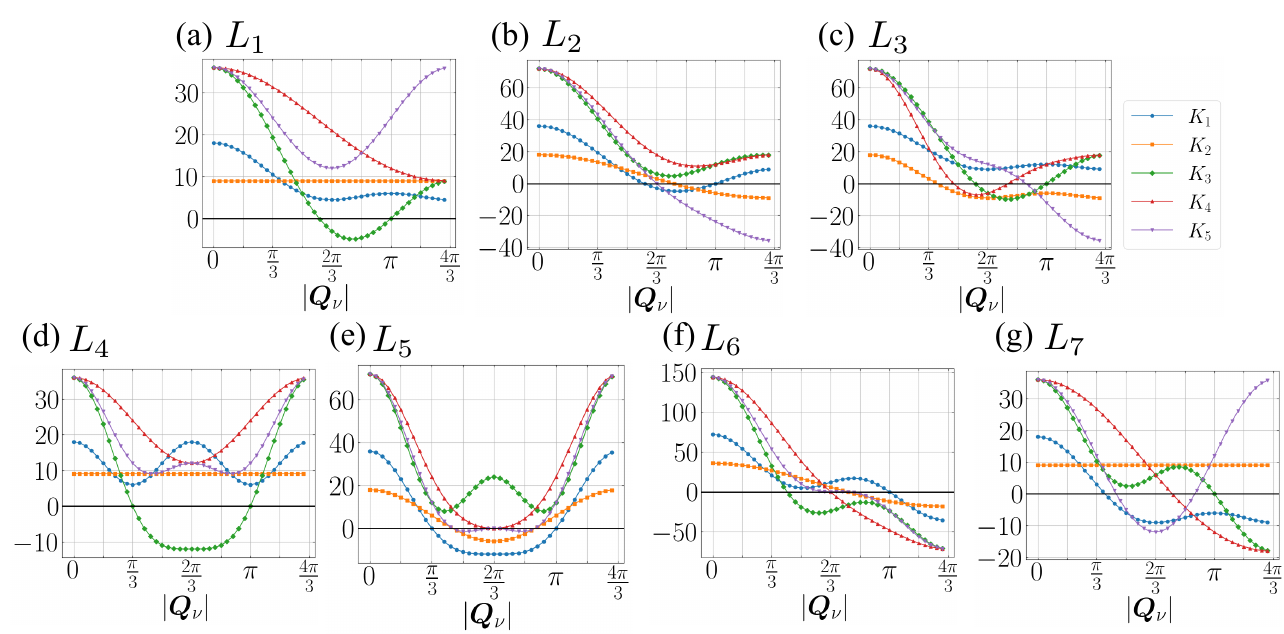}
    \caption{
    %(Color online)
    Momentum-space couplings $K_{\beta}$ generated by each real-space four-spin interaction term $L_\alpha$ ($\alpha=1,\ldots,7$) as a function of the ordering wave vector magnitude $|\bm{Q}_{\nu}|$, assuming $L_\alpha=1$ for each respective term.
    Each panel (a) to (g) corresponds to the contributions from $L_1$ to $L_7$, respectively.
    \label{fig: Fourier transform of L}
    }
  \end{center}
\end{figure*}

To obtain the effective couplings $K_{\beta}$ generated by each $L_\alpha$, we substitute the specific combinations of wave vectors $(\bm{q}, \bm{q}', \bm{q}'')$ listed in Table~\ref{table: allowed set of wave vectors} into the corresponding expression for $K_{\bm{q},\bm{q}',\bm{q}''}^{\alpha}$. 
The results of this calculation are plotted in Fig.~\ref{fig: Fourier transform of L}, which shows the values of the momentum-space couplings as a function of the ordering wave vector magnitude $|\bm{Q}_\nu|$, assuming $L_\alpha=1$ for each respective term.

This analysis reveals several key features. 
First, it is evident that each real-space interaction $L_\alpha$ generates a distinct $|\bm{Q}_{\nu}|$-dependence for the various $K_{\beta}$ couplings. 
This rich and varied behavior demonstrates that a wide range of effective four-spin interaction characteristics can be engineered through a linear combination of these seven microscopic terms. 
Furthermore, we observe that the real-space biquadratic interactions, specifically $L_1$ and $L_4$, predominantly result in a positive momentum-space biquadratic coupling ($K_1 > 0$). 
This is consistent with our earlier interpretation, confirming that these terms tend to suppress single-mode ordering by penalizing large-amplitude spin components and enhance the stability of multiple-$Q$ states as discussed in the previous research~\cite{Hayami_PhysRevB.95.224424, Kartsev2020-biquadratic, Ni2021-biquadratic, Hatanaka2024-biquadratic}.

\subsection{Parameterization of four-body interactions}

Based on the analysis above, we propose a systematic method to parameterize the real-space four-spin interactions. 
This establishes an \textit{inverse design} approach to effectively stabilize desired complex spin textures, as our previous work on bilinear interactions~\cite{Okigami2024-cp}.
As we have seen, each effective momentum-space coupling $K_{\beta}$ is a linear combination of the contributions from the real-space couplings $\{L_\alpha\}$. 
This allows us to frame an inverse problem: instead of calculating the resulting $K_{\beta}$ from a given set of $L_\alpha$, we can specify a set of target $K_{\beta}$ values desired for stabilizing a particular phase and then solve for the microscopic $L_\alpha$ required to produce them.

This linear relation can be expressed concisely in matrix form. 
Let $\bm{K}$ be the vector of target momentum-space couplings and $\bm{L}$ be the vector of the real-space couplings we wish to determine. 
The relationship is then
\begin{equation}
    \bm{K} = M \bm{L},
\end{equation}
where the elements of the transformation matrix, $M_{\beta \alpha}$, are the contributions of each $L_\alpha=1$ to the coupling $K_{\beta}$ (i.e., the values plotted in Fig.~\ref{fig: Fourier transform of L}). Depending on the number of target constraints on $K_{\beta}$ (denoted as $m$) relative to the number of available parameters $L_\alpha$ ($n$), we face three distinct scenarios:

\begin{itemize}
    \item When the number of target $K_{\beta}$ values equals the number of $L_\alpha$ parameters ($m=n$), the system is exactly determined. 
    The problem reduces to solving a standard system of linear equations, and a unique solution for $\bm{L}$ can be readily found by inverting the matrix: $\bm{L} = M^{-1} \bm{K}$.

    \item If we impose more constraints than available parameters ($m > n$), the system is \textit{overdetermined}. 
    An exact solution may not exist, but an optimal set of parameters $\bm{L}$ that best approximates the target $\bm{K}$ can be found using methods such as linear regression.

    \item Conversely, if the number of constraints is less than the number of $L_\alpha$ parameters ($m < n$), the system is \textit{underdetermined}, yielding an infinite number of possible solutions for $\bm{L}$. 
    To select a physically meaningful result from this degenerate solution space, one can employ methods, for instance, the  sparse modeling, to identify a physically efficient solution, such as the one with the fewest non-zero $L_\alpha$ terms.
\end{itemize}

This systematic approach provides a powerful framework for engineering microscopic Hamiltonians designed to stabilize specific, and often elusive, magnetic ground states. 
By leveraging the relationships between real-space and momentum-space couplings, we can gain insights into the types of interactions needed to achieve desired spin textures.

\section{Model Construction with four-spin interactions between Nearest-Neighbor Bonds}

In this section, we demonstrate the practical application of our inverse design approach by constructing a specific spin model that stabilizes the SkX2 phase. 
We begin with a frustrated Heisenberg model on a triangular lattice, which inherently favors a spiral ground state. 
By introducing carefully chosen four-spin interactions, we aim to stabilize the SkX2 phase as the ground state. 
We then validate the stability of the SkX2 phase through comprehensive SA.

\subsection{Model}

As our starting point, we consider a frustrated Heisenberg model with first and third nearest-neighbor interactions on a triangular lattice, with the Hamiltonian given by
\begin{equation}
  \label{eq:H0}
  H_0 = J_1 \sum_{\langle i,j \rangle} \bm{S}_i \cdot \bm{S}_j + J_3 \sum_{\langle\langle\langle i,j \rangle\rangle\rangle} \bm{S}_i \cdot \bm{S}_j,
\end{equation}
where $\langle i,j \rangle$ and $\langle\langle\langle i,j \rangle\rangle\rangle$ denote sums over first- and third-nearest-neighbor pairs, respectively.
By setting the exchange couplings to $J_1 = -1$ (ferromagnetic) and $J_3 = 0.3943$ (antiferromagnetic), the ground state of this model is a degenerate spiral phase. 
This is understood by expressing the Hamiltonian in momentum space as 
\begin{equation}
  H_0 = \frac{1}{2} \sum_{\bm{q}} J(\bm{q}) \bm{S}_{\bm{q}} \cdot \bm{S}_{-\bm{q}},
\end{equation}
with the momentum-space bilinear interaction given by 
\begin{equation}
  \begin{aligned}
    \label{eq:J(q)}
    J(\bm{q}) &= 2J_1 \left( \cos q_x + 2\cos\frac{q_x}{2} \cos\frac{\sqrt{3}q_y}{2} \right) \\
    &+ 2J_3 \left(\cos 2q_x + 2\cos q_x \cos\sqrt{3} q_y \right).
  \end{aligned}
\end{equation}
For the chosen parameters, $J(\bm{q})$ has degenerate minima at the characteristic wave vectors $\bm{Q}_1 = (\pi/3, 0)$ and its symmetry equivalent wave vectors $\bm{Q}_2$ and $\bm{Q}_3$ in terms of the threefold rotation.

To this Hamiltonian, we introduce a four-spin interaction, $H_4$, as a source of multiple-$Q$ instability. 
For simplicity, we restrict our consideration to the three interaction terms constructed exclusively from nearest-neighbor bonds: $L_1$, $L_2$, and $L_3$.

Using our inverse design approach, we set a target for the effective momentum-space couplings that lead to forming an SkX2 state. 
We aim for the following relations at the characteristic wave vector magnitude $|\bm{Q}_\nu|$:
\begin{align*}
    K_1 &= 0.35 |J(\bm{Q}_\nu)|, \\
    K_2 &= -0.70 |J(\bm{Q}_\nu)|, \\
    K_3 &= 0.
\end{align*}
These targets are chosen based on our earlier analysis of the roles of each $K_{\beta}$ in stabilizing the SkX2 phase. 
Our parametrization thus yields a positive $K_1$, which favors a multiple-$Q$ state over a single-$Q$ spiral, and a negative $K_2$, which selectively lowers the energy of sinusoidal superpositions like the SkX2. 
We set $K_3=0$ to isolate the effects of $K_1$ and $K_2$. 
The remaining couplings, $K_4$ and $K_5$, can be ignored in this context as they have no energetic contribution to both the SkX2 and the spiral ansatz and do not influence its stability. 

With three target conditions for our three unknown real-space parameters, the system is exactly determined. 
Solving the corresponding system of linear equations yields a unique set of $\{L_1, L_2, L_3\}$ that realizes the desired momentum-space interactions. 
The resulting parameters are $L_1 = 0.4071$, $L_2 = -0.5748$, and $L_3 = 0.3951$, and we use these values in our subsequent simulations. 
The magnitude of these parameters is considered reasonable, as it has been pointed out that four-spin interactions can be substantial in two-dimensional systems~\cite{Ni2021-biquadratic}.

\subsection{Numerical Results}

To identify the ground state of our constructed model under the external magnetic field, given by
\begin{equation}
  H = H_0 + H_4 - H \sum_i S_i^z,
\end{equation}
we employed a two-step numerical approach utilizing Monte Carlo (MC) simulations and SA. 
Initially, to avoid being trapped in local energy minima, we performed parallel-tempering MC simulations \cite{hukushima1996exchange} with the standard Metropolis algorithm. 
Each replica was evolved for $10^6$ to $5 \times 10^6$ MC sweeps, a duration sufficient for the system to reach thermal equilibrium. 
Following this thermalization, we initiated a SA process. 
Starting with the spin configurations obtained at the lowest MC temperature ($T = 0.1$), we gradually annealed the system down to a final temperature of $T = 0.01$ over an additional $10^6$ MC sweeps, then performed another $10^6$ MC sweeps for measurements. 
We conducted these simulations on triangular lattices of sizes $N = L^2$ with $L = 24$ and $48$, ensuring that the ordering wave vectors $\bm{Q}_1$, $\bm{Q}_2$, and $\bm{Q}_3$ are commensurate with the lattice periodicity, under the periodic boundary conditions.
Hereafter, we show the results for $L=24$, as they are qualitatively consistent with those for $L=48$.

\begin{figure}[bt!]
\begin{center}
\includegraphics[width=0.99\hsize]{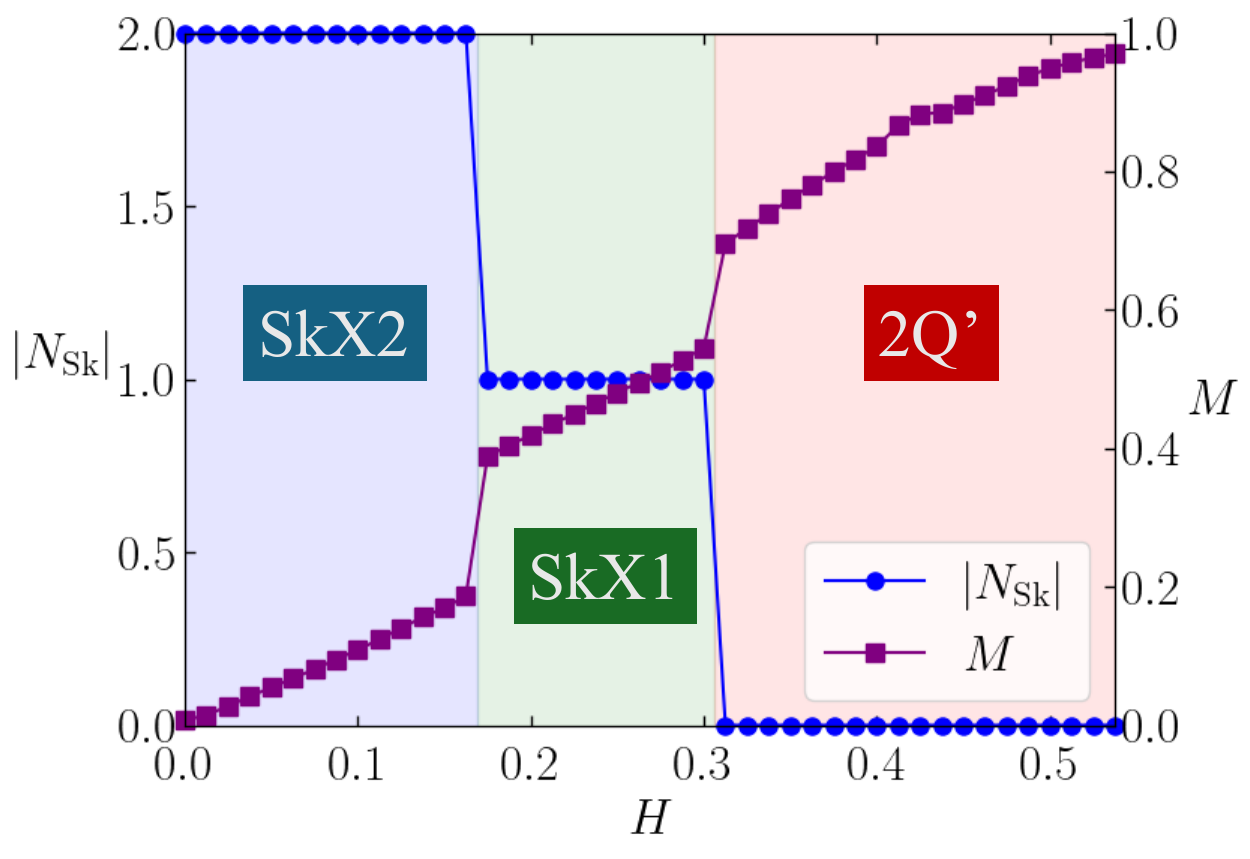}
\caption{
%(Color online)
  Phase diagram showing the skyrmion number $N_{\rm Sk}$ and the magnetization $M$ as a function of the external magnetic field $H$ on the Hamiltonian with four-spin interactions between the nearest-neighbor bonds.
  \label{fig:phase diagram NN bonds}
}
\end{center}
\end{figure}

\begin{figure}[bt!]
\begin{center}
\includegraphics[width=0.99\hsize]{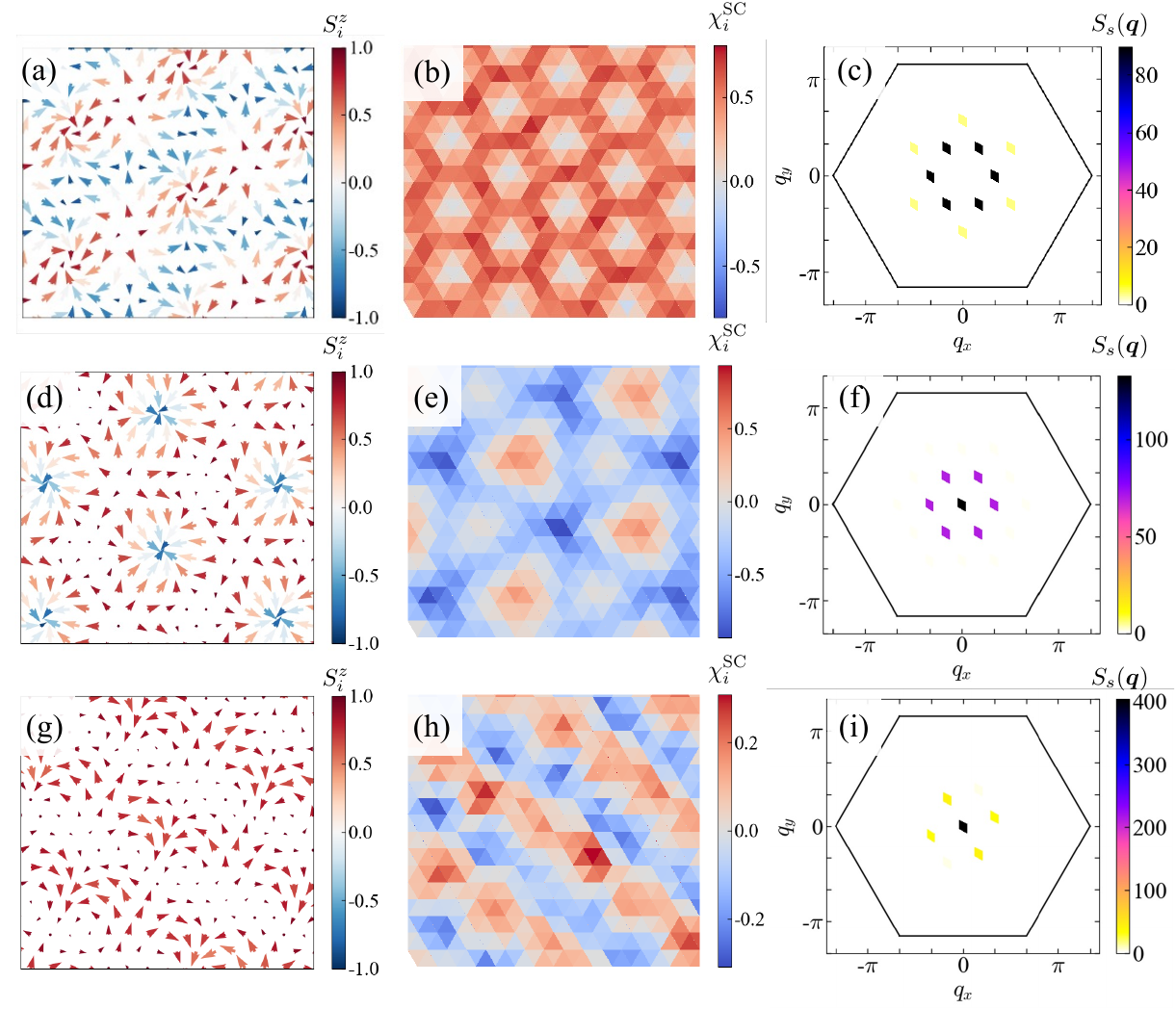}
\caption{
%(Color online)
  Real-space spin configurations (left), local scalar spin chirality (middle), and spin structure factor (right) for (a)-(c) SkX2 at $H=0$, (d)-(f) SkX1 at $H=0.25$, and (g)-(i) 2$Q$' at $H=0.4$.
  \label{fig:configurations NN bonds}
}
\end{center}
\end{figure}

The results of our simulations are summarized in the magnetic field dependent phase diagram shown in Fig.~\ref{fig:phase diagram NN bonds}. 
This figure shows the skyrmion number, $N_{\rm Sk}$, and the magnetization parallel to the field, $M$, as a function of $H$.

At $H=0$, we confirm that the ground state is the SkX2 phase, as was intended by our model construction. 
This is elucidated by the quantized value of $N_{\rm Sk}=2$ and a vanishing net magnetization. 
The detailed characteristics of this phase are presented in Fig.~\ref{fig:configurations NN bonds}. 
Although the SkX2 nature may not be immediately obvious from the real-space spin configuration alone in Fig.~\ref{fig:configurations NN bonds}(a), the plot of the local scalar spin chirality, $\chi^{\rm SC}_i = \bm{S}_i \cdot (\bm{S}_j \times \bm{S}_k)$, reveals a clear triangular lattice pattern with a non-zero net value in Fig.~\ref{fig:configurations NN bonds}(b). 
Furthermore, the spin structure factor $S_s(\bm{q}) = \frac{1}{N} \sum_{\alpha=x,y,z} \sum_{i,j} \expval{S^{\alpha}_i S^{\alpha}_j} e^{i \bm{q} \cdot (\bm{r}_i - \bm{r}_j)}$ in Fig.~\ref{fig:configurations NN bonds}(c) confirms a triple-$Q$ ordering. 
Crucially, the absence of a peak at $\bm{q}=0$ indicates that this SkX2 structure does not possess spontaneous magnetization.

Upon applying an external magnetic field, the system undergoes a transition into the SkX1.
Spin configurations, local scalar spin chirality, and the spin structure factor for this phase are shown in Figs.~\ref{fig:configurations NN bonds}(d)-(f).
In this phase, vortex-like structures are visible in the real-space spin configuration, and the net scalar spin chirality remains non-zero. 
In contrast to the zero-field SkX2 phase, the spin structure factor for the field-stabilized SkX1 phase exhibits a prominent peak at $\bm{q}=\bm{0}$, corresponding to the field-induced uniform magnetization.

With a larger magnetic field, the system transitions into a 2$Q$' phase. 
The same plots for this phase are shown in Figs.~\ref{fig:configurations NN bonds}(g)-(i) as other phases. 
Here, the out-of-plane spin components are largely polarized along the field direction, while the in-plane components retain a periodic modulation, as evidenced by weak peaks at finite $\bm{Q}$ in the spin structure factor. 
The local scalar spin chirality in this 2$Q$' phase averages to zero across the system. 
It is noted that the characteristic wave vectors of this 2$Q$' phase are slightly displaced from the high-symmetry points of the parent spiral state. 
As shown in Fig.~\ref{fig:configurations NN bonds}(i), the spin structure factor at $\bm{q}=\bm{0}$ is considerably large.
Therefore, momentum-space interactions including $\bm{S}_{\bm{0}}$ may potentially have a significant impact, although they have been neglected in the present four-spin interactions analysis.

\section{Ring-Exchange Interaction}

In this section, we demonstrate that the well-known ring-exchange interaction can also stabilize the SkX2 phase when combined with a frustrated Heisenberg model. 
While the ring-exchange term alone is known to favor a four-sublattice all-in/all-out state~\cite{Momoi_PhysRevLett.79.2081}, its effect in the presence of competing interactions is generally not obvious from its real-space form alone. 
By analyzing this interaction through our momentum-space framework, we can predict and confirm its ability to generate the SkX2 phase.

\subsection{Model and Momentum-Space Analysis}

We consider a model where the four-spin ring-exchange interaction is added to the frustrated Heisenberg Hamiltonian in Eq.~(\ref{eq:H0}). 
We focus solely on the four-body terms of the ring-exchange interaction, $H_4^{\rm ring}$, given by
\begin{equation}
  \begin{aligned}
    H_4^{\rm ring} = \sum_{\langle i,j,k,l \rangle} 
    \Big[ &(\bm{S}_i \cdot \bm{S}_j)(\bm{S}_k \cdot \bm{S}_l) + (\bm{S}_i \cdot \bm{S}_l)(\bm{S}_j \cdot \bm{S}_k) \\
    &- (\bm{S}_i \cdot \bm{S}_k)(\bm{S}_j \cdot \bm{S}_l) \Big],
  \end{aligned}
\end{equation}
where the sum is over all minimum four-site plaquettes of the triangular lattice, with sites labeled in a clockwise manner. 
This interaction can be expressed as a linear combination of the real-space four-spin interactions specifically as
\begin{equation}
  \begin{aligned}
    H_4^{\rm ring} &= L_3 \sum_{\langle i,j,k,l \rangle_3} (\bm{S}_i \cdot \bm{S}_j)(\bm{S}_k \cdot \bm{S}_l) \\
    &- L_7 \sum_{\langle i,j,k,l \rangle_7} (\bm{S}_i \cdot \bm{S}_j)(\bm{S}_k \cdot \bm{S}_l),
  \end{aligned}
\end{equation}
where $L_3=1$ and $L_7=1$.
Here, $\langle i,j,k,l \rangle_3$ and $\langle i,j,k,l \rangle_7$ denote sums over all four-site plaquettes corresponding to the interaction types $L_3$ and $L_7$, respectively, as defined in Fig.~\ref{fig: real-space four-body interactions}.
We introduce a coupling constant $J_{\rm ring}$ to control the strength of this interaction, and the total Hamiltonian is
\begin{equation}
  \label{eq:H with ring exchange}
  H = H_0 + J_{\rm ring} H_4^{\rm ring} - H \sum_i S_i^z,
\end{equation}
assuming that any effective bilinear terms arising from the ring-exchange interaction are already absorbed into the parameters of $H_0$.

\begin{figure}[bt!]
\begin{center}
\includegraphics[width=0.99\hsize]{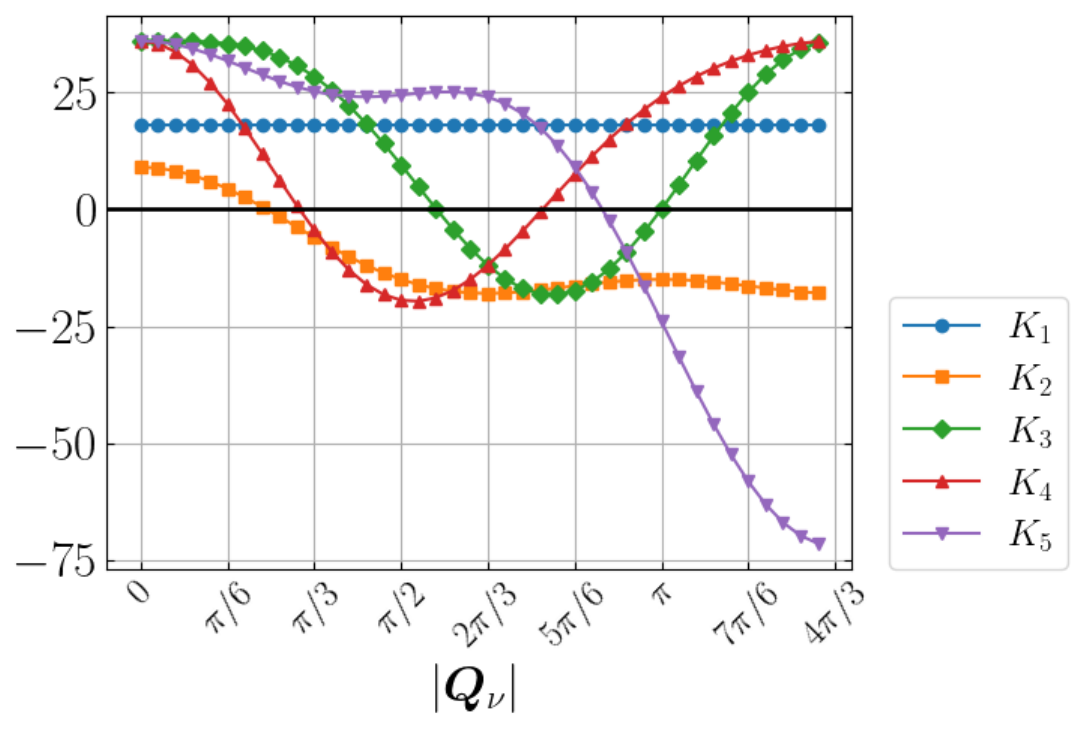}
\caption{
  %(Color online)
  Momentum-space couplings $K_{\beta}$ generated by the ring-exchange interaction as a function of the ordering wave vector magnitude $|\bm{Q}_{\nu}|$, assuming $J_{\rm ring}=1$.
  \label{fig:K ring exchange}
}
\end{center}
\end{figure}

To understand the influence of $H_4^{\rm ring}$ on the degenerate spiral states of the model, we project it onto the momentum-space basis and calculate the effective couplings $\{K_{\beta}\}$. 
The results are plotted as a function of the ordering wave vector magnitude in Fig.~\ref{fig:K ring exchange}. 
For the specific wave vector of interest in $H_0$, $|\bm{Q}_{\nu}|=\pi/3$, the figure clearly shows that the ring-exchange interaction yields $K_1>0$ and $K_2<0$. 
This is precisely the condition we identified as being favorable for the SkX2 phase: a positive $K_1$ favors multiple-$Q$ states, and a negative $K_2$ selectively stabilizes the sinusoidal superposition.

\subsection{Numerical Results}

We performed simulations on the combined Hamiltonian in Eq.~(\ref{eq:H with ring exchange}), with the ring-exchange coupling set to $J_{\rm ring}=0.1$. 
The simulation protocol is identical to that described in the previous section.

\begin{figure}[bt!]
\begin{center}
\includegraphics[width=0.99\hsize]{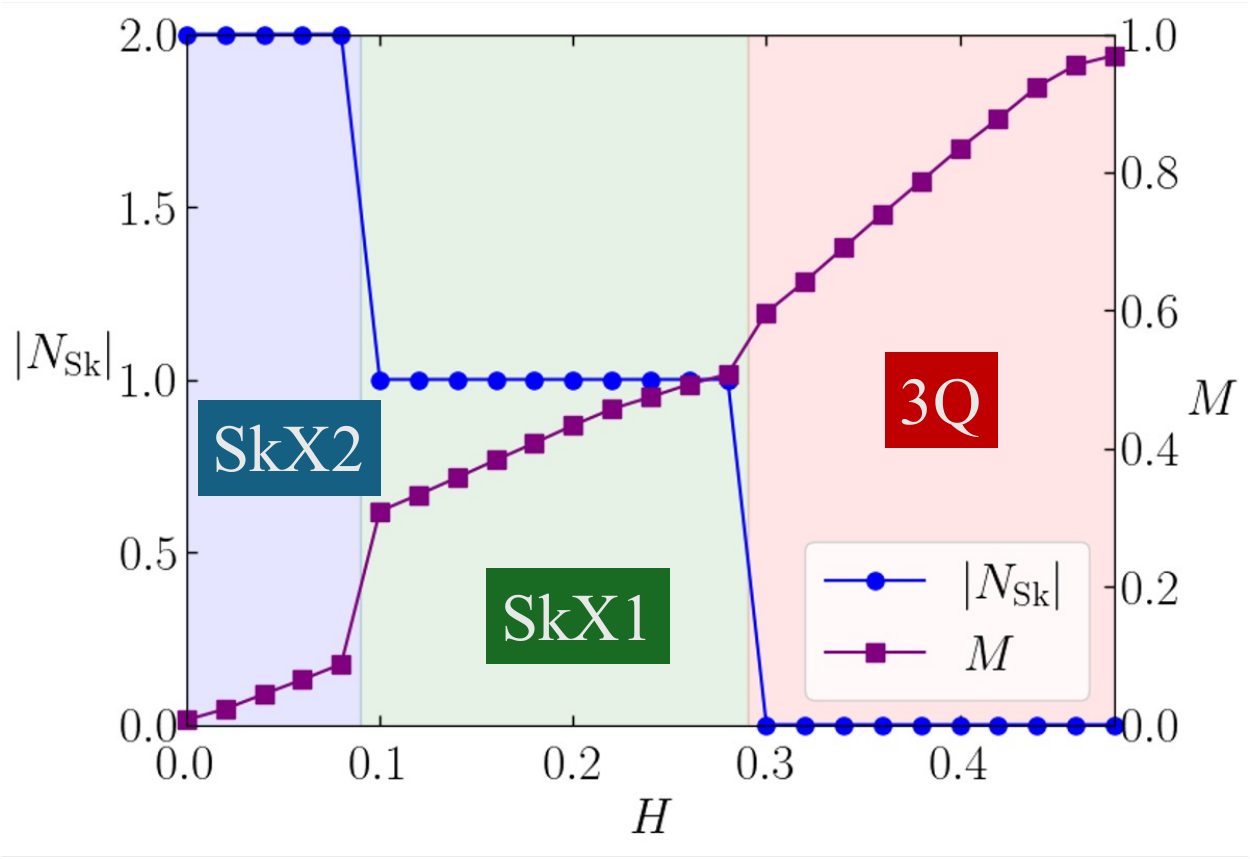}
\caption{
  %(Color online)
  Phase diagram showing the skyrmion number $N_{\rm Sk}$ and the magnetization $M$ as a function of the external magnetic field $H$ on the Hamiltonian with ring-exchange interaction.
  \label{fig:phase diagram ring}
}
\end{center}
\end{figure}

The resulting magnetic phase diagram is shown in Fig.~\ref{fig:phase diagram ring}. 
The phase sequence is broadly similar to that found with our previous model, with one notable difference in the high-field region, where a non-skyrmionic 3$Q$ phase emerges.

\begin{figure}[bt!]
\begin{center}
\includegraphics[width=0.99\hsize]{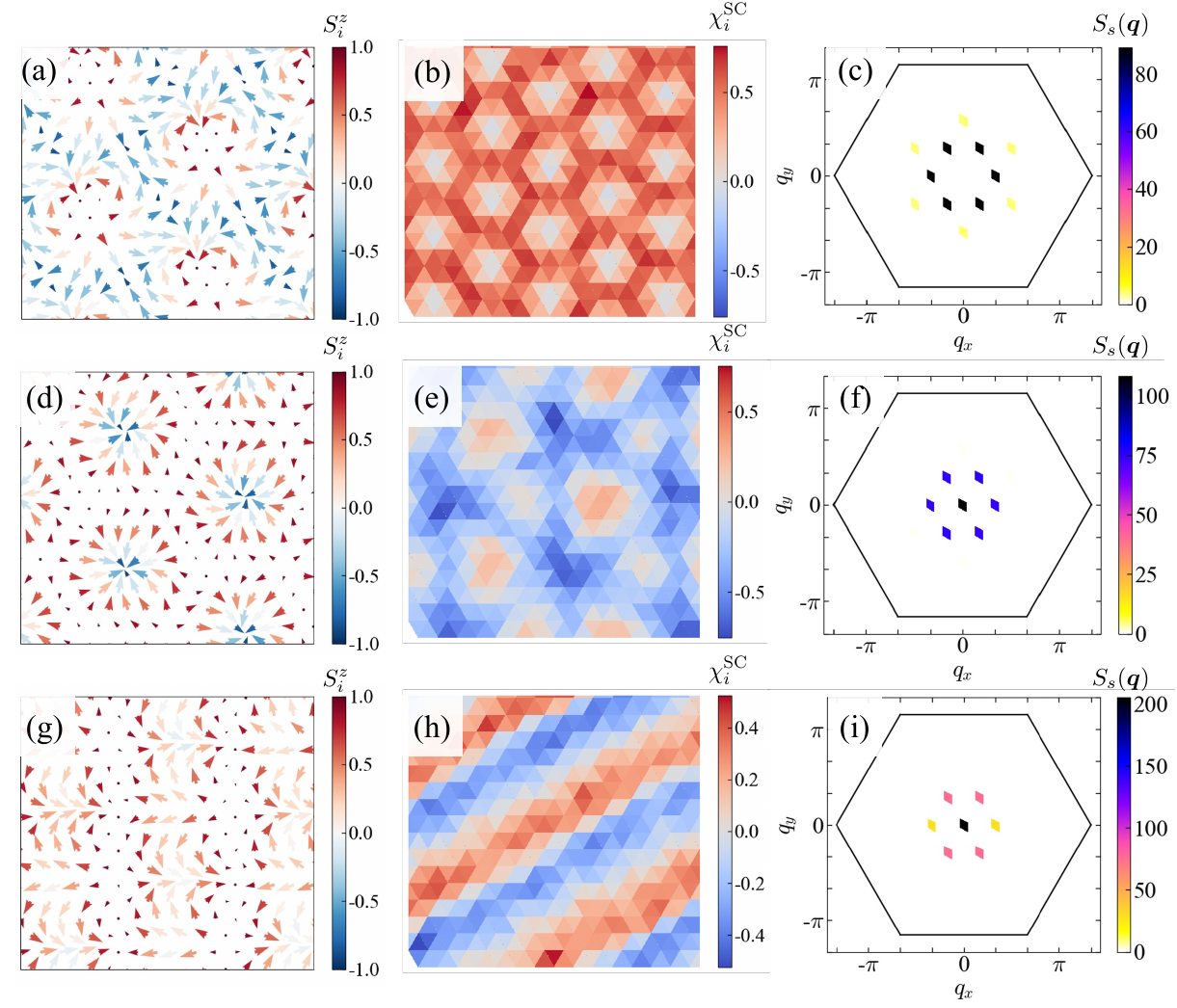}
\caption{
  %(Color online)
  Real-space spin configurations (left), local scalar spin chirality (middle), and spin structure factor (right) for (a)-(c) SkX2 at $H=0$, (d)-(f) SkX1 at $H=0.2$, and (g)-(i) 3$Q$ at $H=0.3$.
  \label{fig:configurations ring}
}
\end{center}
\end{figure}

At zero magnetic field, the system successfully stabilizes the SkX2 phase, confirmed by $N_{\rm Sk}=2$. 
The detailed spin texture, the local scalar spin chirality, and the spin structure factor are shown in Figs.~\ref{fig:configurations ring}(a)-(c). 
While the real-space spin configuration in Fig.~\ref{fig:configurations ring}(a) appears to have more of a vortex-like character than the solution in the previous section, this is a superficial difference. 
At $H=0$, the system possesses global SO(3) spin-rotational symmetry, meaning that the two spin textures are physically equivalent and can be transformed into one another by a uniform spin rotation. 
This can be understood by examining the common features present in the triangular lattice of scalar spin chirality and the triple-$Q$ spin structure factor with no $\bm{q} = \bm{0}$ peak. 
These features are analogous to those observed in the previous case, thereby unambiguously identifying the phase as the SkX2.

As the magnetic field increases, the system transitions into the SkX1 phase with $N_{\rm Sk}=1$, which has the same characteristics as the SkX1 phase discussed previously. 
This phase is detailed in Figs.~\ref{fig:configurations ring}(d)-(f). 
At even higher fields, we find a distinct non-skyrmionic triple-$Q$ phase (3$Q$), shown in Figs.~\ref{fig:configurations ring}(g)-(i). 
This phase is characterized by $N_{\rm Sk}=0$, a vanishing net scalar spin chirality, and unequal peak intensities in its spin structure factor. 
Specifically, the structure factor exhibits two in-plane peaks, one finite-$Q$ out-of-plane peak, and a field-induced peak at $\bm{q}=\bm{0}$. 
We note that the ordering wave vectors in this phase do not appear to be shifted from the high-symmetry points, indicating that the locking tendency of the ordering wave vector is stronger than that in the previous case in Sec.~III.

\section{Discussion}

While our present work has focused on the stabilization of the SkX2 phase, the inverse design methodology we have developed is broadly applicable. 
The framework connecting microscopic real-space interactions $\{L_\alpha\}$ to their effective momentum-space couplings $\{K_{\beta}\}$ provides a powerful tool for engineering a wide variety of complex spin textures, such as those formed by skyrmioniums or hopfions, which currently lack well-established microscopic models that guarantee their stability as the ground state. 
Recently, we have demonstrated that four-spin interactions can induce a broken helix state in a one-dimensional spin chain~\cite{Gen2025-brokenhelix, HayamiOkigami2025_brokenhelix}, which is a multiple-$Q$ state composed of two symmetry-unrelated wave vectors.
Our approach offers a systematic pathway to construct such models by targeting the specific momentum-space interactions that would favor these desired topological spin structures.

Our framework provides a powerful tool for exploring complex magnetism from two complementary perspectives. 
The first is a top-down or inverse design approach, which focuses on constructing a model to realize a specific, desired spin structure. 
In this direction, one first identifies the effective momentum-space interactions $\{K_{\beta}\}$ required to stabilize the target phase. 
The next step is to solve for the microscopic real-space couplings $\{L_\alpha\}$ that will generate these interactions. 
This is the strategy we employed in our initial example, where we successfully engineered an SkX2-stabilizing model using only nearest-neighbor four-spin terms.
Such a strategy can also be applied to infer the underlying microscopic spin model that reproduces the experimentally observed phase diagram, including multiple-$Q$ topological spin textures beyond phenomenological momentum-space models. 
This approach is particularly relevant to materials where four-spin interactions are believed to play a crucial role, such as EuPtSi~\cite{hayami2021field}.

The second perspective is a bottom-up, or material-driven, analysis. 
When the symmetry-allowed multi-body interactions for a given material or model are known, our framework can be used to calculate the resulting momentum-space couplings $\{K_{\beta}\}$. 
The character of these couplings then provides a clear indicator of which complex spin textures the system is likely to favor. 
Our successful stabilization of the SkX2 phase using the ring-exchange interaction exemplifies this direction; by analyzing its momentum-space signature, we could predict its tendency to form the SkX2, a conclusion not immediately obvious from its real-space form. 

Furthermore, this framework is not limited to four-body interactions. 
The fundamental concept of mapping microscopic real-space terms to effective momentum-space couplings for a specific multiple-$Q$ ansatz can be readily generalized. 
This framework can be extended to explore the effects of higher-order interactions, such as three-body and other multi-body terms, which may play a crucial role in stabilizing complex magnetic phases. 
In a similar context, the present framework can be extended to crystalline-dependent anisotropic four-body interactions including the chiral biquadratic interaction~\cite{lounis2020multiple, Mankovsky_PhysRevB.101.174401, Paul2020role, Yambe_PhysRevB.106.174437}. 
By systematically analyzing the contributions of these interactions, we can develop a more comprehensive understanding of the rich landscape of possible spin configurations.

Nevertheless, a crucial question remains: how to systematically determine the optimal set of target values $\{K_{\beta}\}$ for an arbitrary, complex spin texture. 
This remains an open problem for the future investigation. 
Should a general method for determining the target $\{K_{\beta}\}$ be developed, it would enable a complete, two-stage design protocol: first, identify the ideal phenomenological couplings for a desired phase, and second, use our framework to construct the microscopic Hamiltonian that produces them.

\section{Summary}

In this paper, we proposed and demonstrated a systematic framework for the inverse design of spin Hamiltonians with higher-order interactions, aimed at stabilizing exotic multiple-$Q$ topological magnetic textures.
While higher-order interactions are known to be crucial for realizing complex spin states, a general principle for constructing models with such terms has been lacking. 
Our approach addresses this challenge by leveraging a momentum-space representation of four-spin interactions. 
This method allows us to reframe the problem: we first identify the phenomenological momentum-space couplings \{$K_{\beta}$\} that favor a target spin texture, and then we systematically solve for the microscopic real-space couplings \{$L_{\alpha}$\} required to produce them.

We validated our methodology through two key demonstrations focusing on stabilizing the SkX2 state. 
First, using the top-down design approach, we successfully constructed a model with only three nearest-neighbor four-spin interactions that stabilizes the SkX2 phase as its ground state, which we confirmed via SA. 
Second, we applied our framework in a bottom-up, predictive manner to the well-known ring-exchange interaction. 
Our analysis revealed that this interaction naturally produces the key momentum-space signatures ($K_1>0$ and $K_2<0$) required for the SkX2 stabilization, a non-trivial insight that was also verified by our numerical simulations.

Our results establish a clear and intuitive pathway for both engineering distinct spin models for target phases and for understanding the role of multi-body interactions in existing models or materials. 
This momentum-space framework overcomes the limitations of traditional methods applicable only to bilinear interactions and provides a versatile tool for the broader exploration of complex magnetism. 
The proposed methodology paves the way for the systematic design and discovery of novel topological spin states stabilized by a rich variety of multi-spin interactions.

\section*{Acknowledgments}
We thank T. Hatanaka for fruitful discussions on biquadratic interactions.
K.O. was supported by JST SPRING, Grant Number JPMJSP2108.
This was also supported by JSPS KAKENHI Grants Numbers JP22H00101, JP22H01183, JP23H04869, JP23K03288, and by JST CREST (JPMJCR23O4) and JST FOREST (JPMJFR2366).

\appendix

\section{Derivation of momentum-space four-body interactions from real-space four-body interactions}
\label{appendix: derivation of K from L}
The contribution from a single real-space term $L_\alpha$ to the general expression for the momentum-space interaction given in Eq.~\ref{eq:K_q_q'_q''} is denoted as $K_{\bm{q},\bm{q}',\bm{q}''}^{\alpha}$. 
The explicit expressions are as follows:
\begin{widetext}
  \begin{equation}
    \begin{aligned}
      K_{\bm{q},\bm{q}',\bm{q}''}^{1} &= L_1 \Big\{
        \cos \left[(\bm{q} + \bm{q}'') \cdot \bm{a}_1 \right] 
        + \cos \left[(\bm{q} + \bm{q}'') \cdot \bm{a}_2 \right] 
        + \cos \left[(\bm{q} + \bm{q}'') \cdot \bm{a}_3 \right] 
        \Big\}, \\
      K_{\bm{q},\bm{q}',\bm{q}''}^{2} &= 2 L_2 \Big\{
        \cos \left[\bm{q}' \cdot \bm{a}_1 + (\bm{q} + \bm{q}' + \bm{q}'') \cdot \bm{a}_2 \right] \\
        &\quad\quad\quad+ \cos \left[\bm{q}' \cdot \bm{a}_2 + (\bm{q} + \bm{q}' + \bm{q}'') \cdot \bm{a}_3 \right] \\
        &\quad\quad\quad+ \cos \left[\bm{q}' \cdot \bm{a}_3 + (\bm{q} + \bm{q}' + \bm{q}'') \cdot \bm{a}_1 \right]
        \Big\}, \\
      K_{\bm{q},\bm{q}',\bm{q}''}^{3} &= L_3 \Big\{
        \cos \left[\bm{q}' \cdot \bm{a}_1 + \bm{q}'' \cdot (\bm{a}_1 - \bm{a}_2) + (\bm{q} + \bm{q}' + \bm{q}'') \cdot \bm{a}_2 \right] \\
        &\quad\quad+ \cos \left[\bm{q}' \cdot \bm{a}_1 + \bm{q}'' \cdot (\bm{a}_1 - \bm{a}_3) + (\bm{q} + \bm{q}' + \bm{q}'') \cdot \bm{a}_3 \right] \\
        &\quad\quad+ \cos \left[\bm{q}' \cdot \bm{a}_2 + \bm{q}'' \cdot (\bm{a}_2 - \bm{a}_3) + (\bm{q} + \bm{q}' + \bm{q}'') \cdot \bm{a}_3 \right] \\
        &\quad\quad+ \cos \left[\bm{q}' \cdot \bm{a}_2 + \bm{q}'' \cdot (\bm{a}_2 -\bm{a}_1) + (\bm{q} + \bm{q}' + \bm{q}'') \cdot \bm{a}_1 \right] \\
        &\quad\quad+ \cos \left[\bm{q}' \cdot \bm{a}_3 + \bm{q}'' \cdot (\bm{a}_3 - \bm{a}_1) + (\bm{q} + \bm{q}' + \bm{q}'') \cdot \bm{a}_1 \right] \\
        &\quad\quad+ \cos \left[\bm{q}' \cdot \bm{a}_3 + \bm{q}'' \cdot (\bm{a}_3 - \bm{a}_2) + (\bm{q} + \bm{q}' + \bm{q}'') \cdot \bm{a}_2 \right]
        \Big\}, \\
      K_{\bm{q},\bm{q}',\bm{q}''}^{4} &= L_4 \Big\{
        \cos \left[(\bm{q} + \bm{q}'') \cdot (\bm{a}_1 - \bm{a}_2) \right] 
        + \cos \left[(\bm{q} + \bm{q}'') \cdot ( \bm{a}_2 - \bm{a}_3) \right] 
        + \cos \left[(\bm{q} + \bm{q}'') \cdot (\bm{a}_3 - \bm{a}_1) \right] 
        \Big\}, \\
      K_{\bm{q},\bm{q}',\bm{q}''}^{5} &= 2 L_5 \Big\{
        \cos \left[\bm{q}' \cdot (\bm{a}_1 - \bm{a}_2) + (\bm{q} + \bm{q}' + \bm{q}'') \cdot (\bm{a}_3 - \bm{a}_1) \right] \\
        &\quad\quad\quad + \cos \left[\bm{q}' \cdot (\bm{a}_2 - \bm{a}_3) + (\bm{q} + \bm{q}' + \bm{q}'') \cdot (\bm{a}_1 - \bm{a}_2) \right] \\
        &\quad\quad\quad + \cos \left[\bm{q}' \cdot (\bm{a}_3 - \bm{a}_1) + (\bm{q} + \bm{q}' + \bm{q}'') \cdot (\bm{a}_2 - \bm{a}_3) \right]
        \Big\}, \\
      K_{\bm{q},\bm{q}',\bm{q}''}^{6} &= 2 L_6 \Big\{
        \cos \left[\bm{q}' \cdot \bm{a}_1 + (\bm{q} + \bm{q}' + \bm{q}'') \cdot (\bm{a}_2 - \bm{a}_1) \right] \\
        &\quad\quad\quad+ \cos \left[\bm{q}' \cdot \bm{a}_1 + (\bm{q} + \bm{q}' + \bm{q}'') \cdot (\bm{a}_3 - \bm{a}_1) \right] \\
        &\quad\quad\quad+ \cos \left[\bm{q}' \cdot \bm{a}_2 + (\bm{q} + \bm{q}' + \bm{q}'') \cdot (\bm{a}_3 - \bm{a}_2) \right] \\
        &\quad\quad\quad+ \cos \left[\bm{q}' \cdot \bm{a}_2 + (\bm{q} + \bm{q}' + \bm{q}'') \cdot (\bm{a}_1 - \bm{a}_2) \right] \\
        &\quad\quad\quad+ \cos \left[\bm{q}' \cdot \bm{a}_3 + (\bm{q} + \bm{q}' + \bm{q}'') \cdot (\bm{a}_1 - \bm{a}_3) \right] \\
        &\quad\quad\quad+ \cos \left[\bm{q}' \cdot \bm{a}_3 + (\bm{q} + \bm{q}' + \bm{q}'') \cdot (\bm{a}_2 - \bm{a}_3) \right]
        \Big\}, \\
      K_{\bm{q},\bm{q}',\bm{q}''}^{7} &= L_7 \Big\{
        \cos \left[\bm{q}' \cdot (\bm{a}_1 - \bm{a}_2) + \bm{q}'' \cdot \bm{a}_1 + (\bm{q} + \bm{q}' + \bm{q}'') \cdot \bm{a}_2 \right] \\
        &\quad\quad\quad+ \cos \left[\bm{q}' \cdot (\bm{a}_2 - \bm{a}_3) + \bm{q}'' \cdot \bm{a}_2 + (\bm{q} + \bm{q}' + \bm{q}'') \cdot \bm{a}_3 \right] \\
        &\quad\quad\quad+ \cos \left[\bm{q}' \cdot (\bm{a}_3 - \bm{a}_1) + \bm{q}'' \cdot \bm{a}_3 + (\bm{q} + \bm{q}' + \bm{q}'') \cdot \bm{a}_1 \right]
        \Big\}, \\
    \end{aligned}
  \end{equation}
\end{widetext}
where $\bm{a}_1 = (1,0)$, $\bm{a}_2 = (-1/2, \sqrt{3}/2)$, and $\bm{a}_3 = -\bm{a}_1 - \bm{a}_2$ are the primitive lattice vectors of the triangular lattice.

\newpage

\bibliography{ref}

\end{document}